\documentclass{jaa}
\usepackage{natbib}
\usepackage{multicol}
\usepackage{caption}
\usepackage{graphicx}
\usepackage{multirow}
\usepackage{subfigure}
\usepackage{xcolor}
\usepackage{adjustbox}
\usepackage[colorlinks=true,linkcolor=blue,citecolor=blue]{hyperref}
\begin{document}

\title{ALMA detection of hydrogen cyanide (HCN) in the atmosphere of Saturn}

\author{Arijit Manna\textsuperscript{1*}, Sabyasachi Pal\textsuperscript{1}}
\affilOne{\textsuperscript{1}Department of Physics and Astronomy, Midnapore City College, Paschim Medinipur, West Bengal, India 721129\\}
 
\twocolumn[{
\maketitle
\corres{amanna.astro@gmail.com}

\msinfo{--}{--}

\vspace{0.5cm}
\begin{abstract}
In the planetary atmosphere, hydrogen cyanide (HCN) is an important nitrogen (N)-bearing molecule that plays a key role in the formation of several biomolecules via chain reactions. The presence of HCN characterizes the stratospheric composition of the solar gas planets and exoplanets. For several years, many observations have failed to identify the rotational and vibrational emission lines of HCN from the atmosphere of Saturn using ground- and space-based radio telescopes. We present the successful detection of the rotational emission line of HCN from the atmosphere of Saturn using the Atacama Large Millimeter/Submillimeter Array (ALMA) band 7. We detected the J = 4--3 transition line of the HCN from the eastern and western limbs of Saturn with $\geq5\sigma$ statistical significance. The derived abundances of HCN in the western and eastern limbs are 6.19 ppb and 2.90 ppb, respectively. We claim that HCN is formed in the atmosphere of Saturn via the photolysis of methane (CH$_{4}$) and ammonia (NH$_{3}$).
\end{abstract}

}]
\doinum{xyz/123}
\artcitid{\#\#\#\#}
\volnum{000}
\year{2021}
\pgrange{1--11}
\setcounter{page}{1}
\lp{11}

\section{Introduction}
\label{sec:intro}  
In the planetary atmosphere, hydrogen cyanide (HCN) is an important nitrogen (N)-bearing molecule for the synthesis of several complex organic molecules \citep{tok81, ka83}. The HCN molecule contains a --C $\equiv$ N functional group, which plays a major role in the formation of peptides, nucleic acids, amino acids, and both nucleobases, that can be the building blocks of DNA and RNA in the planetary atmosphere \citep{mil57, oro61}. The HCN molecule may play an important role in the formation of adenine ({C$_{5}$H$_{5}$N$_{5}$}) via the neutral-neutral reactions \citep{oro61}. Adenine is one of the nucleobases of DNA and RNA and plays a major role in the origin of life in the possible primitive Earth conditions \citep{oro61}. HCN molecules may also be responsible for the production of chromophores in planetary atmospheres \citep{wo69}. The HCN molecule acts as a potential tracer of disequilibrium processes such as the transport of materials from the deep atmosphere or lightning \citep{le80}. In a planetary atmosphere, HCN is mainly formed via the dissociation of methane (CH$_{4}$) and nitrogen (N$_{2}$) \citep{cat07, per19}. Several energy sources can dissociate CH$_{4}$ and N$_{2}$ in planetary atmospheres, including galactic cosmic rays, lightning, and ultraviolet light \citep{per19, per20}.

In our solar system, Saturn's moon Titan had the most HCN-rich atmosphere because of its high atmospheric C/O ratio \citep{rim19}. Earlier, four instruments aboard the Cassini spacecraft detected the HCN from the lower ($<$600 km) and upper ($>$700 km) atmospheres of Titan with mixing ratios of $\sim$0.1--10 ppm (parts per million) and $\sim$0.1\%--5\% (parts per thousand) \citep{mag09, vin10, ad11, kos11}. The galactic cosmic rays and ultraviolet lights are responsible for dissociating CH$_{4}$ and N$_{2}$ to create radical compounds in the upper and lower atmospheres of Titan \citep{gro09, gro11, vui19}. The emission lines of HCN were also detected in the atmospheres of Neptune and Pluto, with abundances of $\sim$1 ppb (parts per billion) and $\sim$40 ppm, respectively, \citep{mar93, lel17}. \citet{ii20} detected the belt-like distribution of HCN in the atmosphere of Neptune using the ALMA. In the equatorial region and 60$^{\circ}$ S of Neptune, the abundance of HCN is 1.66 ppb and 1.17 ppb, respectively \citep{ii20}. The concentration of HCN in the upper atmosphere of Uranus is 0.1 ppb \citep{mar93}. Previously, \cite{tok81} detected HCN from Jupiter with a mixing ratio of 2$\times$10$^{-9}$.

\begin{figure}
	\centering
	\includegraphics[width=0.45\textwidth]{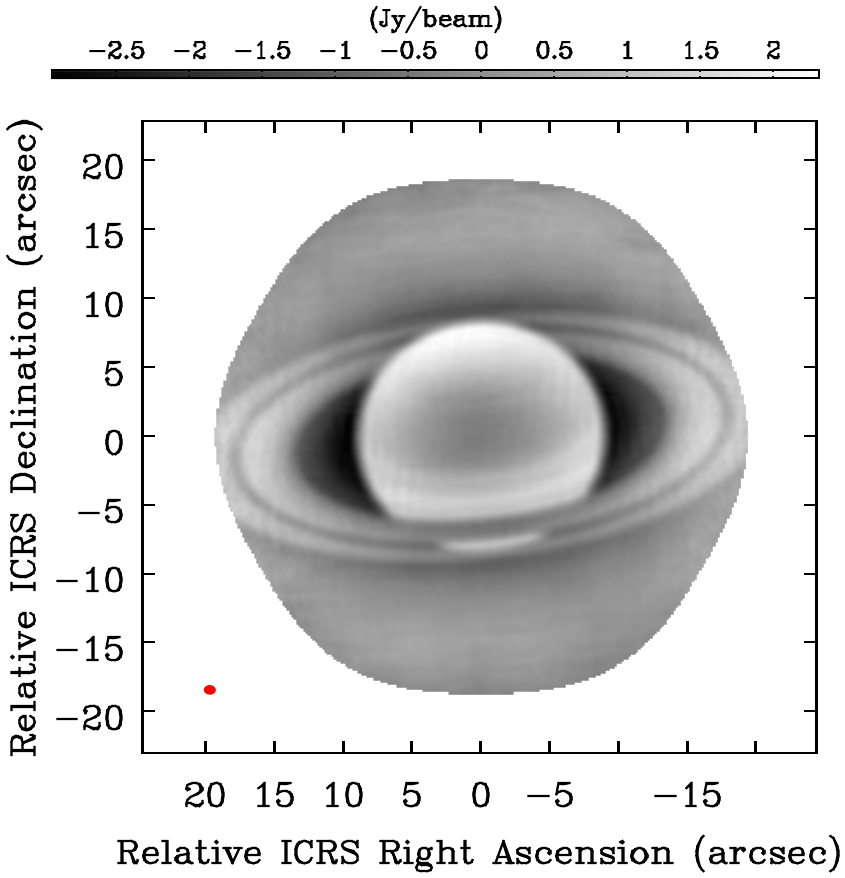}
	\caption{Submillimeter-wavelength continuum emission image of Saturn at a frequency of 350.14 GHz (856.19 $\mu$m). In the image, the red circle indicates the synthesized beam. The corresponding synthesized beam size is 0.78$^{\prime\prime}\times$0.61$^{\prime\prime}$. The RMS of the continuum image is 95.2 mJy beam$^{-1}$.}
	\label{fig:cont}
\end{figure}

In our solar system, the study of HCN in the atmosphere of Saturn at sub-millimeter wavelengths is important because HCN acts as a possible precursor of many major prebiotic molecules \citep{lel84}. In the last several years, scientists have attempted to detect evidence of HCN in the atmosphere of Saturn without success. Earlier, \citet{tok81} searched for evidence of HCN in the atmosphere of Saturn in the infrared wavelength range, but they did not find any evidence of HCN in the atmosphere of Saturn because of the lower signal-to-noise ratio (SNR). \citet{tok81} calculated the upper-limit mixing ratio of HCN in Saturn's atmosphere to be $\leq$5$\times$10$^{-10}$. Earlier, \cite{ka83} suggested that the HCN mixing ratio in the atmosphere at Saturn is 5$\times$10$^{-10}$, which was estimated based on photochemical reactions. According to \citet{lel84}, the submillimeter spectral range is the best for searching for evidence of HCN in Saturn's atmosphere because, using the heterodyne technique, the possibility of HCN line resolving power is at least 10$^{6}$ and the four strongest HCN rotational lines lie between 800 $\mu$m and 3 mm.

In this letter, we present the detection of the rotational emission line of HCN in the atmosphere of Saturn using ALMA. In Section~\ref{sec:obs}, we briefly describe our observations and data reduction. The result and discussion of the detection of HCN are described in Section~\ref{sec:emi}. The conclusion is presented in Section~\ref{sec:con}.

\begin{figure}
	\centering
	\includegraphics[width=0.5\textwidth]{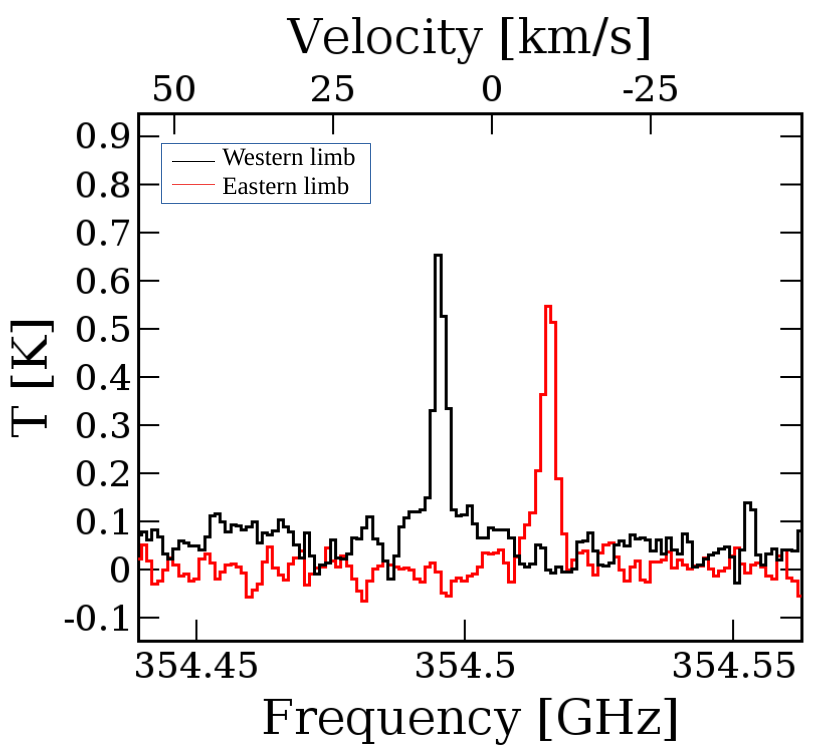}
	\caption{The extracted molecular emission spectrum of HCN (J = 4--3) in Saturn's atmosphere. In the spectrum, the red spectral profile indicates the HCN spectra extracted from the eastern limb, and the black spectral profile indicates the HCN spectra extracted from the western limb. The spectral shifting occurred because of high-velocity stratospheric winds.}
	\label{fig:SPEC}
\end{figure}

\begin{figure*}
\centering
\includegraphics[width=1.0\textwidth]{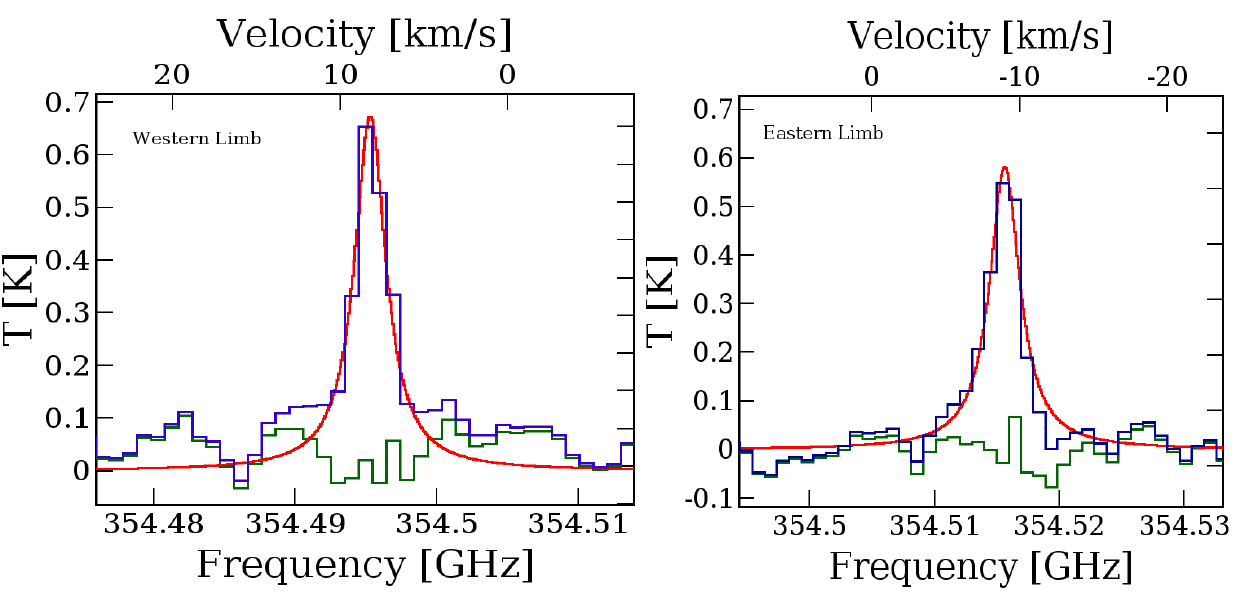}
\caption{Rotational emission line of HCN (J = 4--3) with the best-fit model spectrum obtained from the PSG. The blue spectra indicate the observed spectra of HCN in the western and eastern limbs of Saturn, and the red synthetic spectra indicate the best-fit PSG radiative transfer model with an abundance of 6.19 ppb (western limb) and 2.90 ppb (eastern limb), respectively. The green spectra indicate the residuals of the molecular spectra.}
\label{fig:HCNSPEC}
\end{figure*}

\section{Observation and data reductions}
\label{sec:obs}
We used the archival data of the solar gas planet Saturn, which was observed using the high-resolution Atacama Large Millimeter/Submillimeter Array (ALMA) band 7 with a 12 m array (PI: Fouchet, Thierry). The objective of this observation was to directly measure Saturn's stratospheric winds using the emission lines of hydrogen cyanide (HCN) with transition J = 4--3 and carbon monoxide (CO) with transition J = 3--2. Saturn was observed on May 25, 2018, with the phase centre of ($\alpha,\delta$)$ \rm J2000$ = 18:37:56.412, --22:15:52.617. During the observation, a total of forty-four antennas were used, with an on-source integration time of 476.40 sec. The observed frequency ranges were 340.98--342.96 GHz, 342.98--344.96 GHz, 354.42--354.54 GHz, and 354.98--356.96 GHz, with spectral resolutions of 1128.91 kHz, 70.56 kHz, 282.23 kHz, and 1128.91 kHz, respectively. During the observation, the angular diameter of Saturn was 17.44$^{\prime\prime}$, the geocentric distance was 9.21 AU, and the angular resolution was 0.70$^{\prime\prime}$, 0.78$^{\prime\prime}$, 0.82$^{\prime\prime}$, and 0.85$^{\prime\prime}$ for the corresponding frequency ranges of 340.98--342.96 GHz, 342.98--344.96 GHz, 354.42--354.54 GHz, and 354.98--356.96 GHz, respectively. The minimum and maximum baselines during the observation were 15.0 m and 313.7 m. As a result of the lack of short spacing with the interferometer, the disk flux of Saturn is filtered out, and only the emission from the limbs is preserved. The shortest baseline observation can directly measure Saturn's stratospheric winds between the eastern and western limbs. During the Saturn observation, the flux and bandpass calibrators were taken as J1750+0939, and the phase calibrator was taken as J1832--2039. Previously, \citet{ben22} also used this data, but the authors did not focus on studying the mixing ratio, vertical distribution, and formation pathways of HCN in the atmosphere of Saturn.

After downloading the archival raw data of Saturn from the ALMA archive, we used the Common Astronomy Software Application ({\tt CASA 5.4.1}) with a standard data reduction pipeline delivered by the ALMA observatory for data analysis and imaging \citep{m07}. We imported the ALMA data into CASA using task {\tt IMPORTASDM}. During the data analysis, we used the Perley-Butler 2017 flux calibrator model to match the continuum flux density of the flux calibrator with 5\% accuracy using the task {\tt SETJY} in CASA \citep{pal17}. We applied the CASA pipeline tasks {\tt hifa\_bandpassflag} and {\tt hifa\_flagdata} to remove the bad antenna data. During the bandpass calibration, we carefully removed the telluric effect from the data by flagging the channels where the telluric absorption lines were observed. We do not observe any telluric lines near the lines of CO and HCN. We found a strong telluric absorption line of ozone (O$_{3}$) at a frequency of 355.01 GHz, and we removed that line during the data analysis. After the flux, bandpass, and gain calibration, we separated the target dataset into another dataset using the task {\tt MSTRANSFORM} with all available rest frequencies. After splitting the target data, we used task {\tt TCLEAN} with the {\tt HOGBOM} algorithm to create the continuum image of Saturn. We used the line-free channels during the production of the continuum emission images. After that, we used the CASA task {\tt UVCONTSUB} for continuum subtraction because continuum subtraction is essential for studying the spectral line. After continuum subtraction, we used the task {\tt TCLEAN} with the {\tt SPECMODE = CUBE} parameter for each rest frequency to create the spectral data cubes of Saturn. To correct the primary beam pattern in the synthesized image, we applied the {\tt IMPBCOR} task in the CASA. The data analysis flowchart is shown in Figure {\color{blue}A1}.

\section{Result and discussion}
\label{sec:emi}

\begin{figure*}
	\centering
	\includegraphics[width=1.0\textwidth]{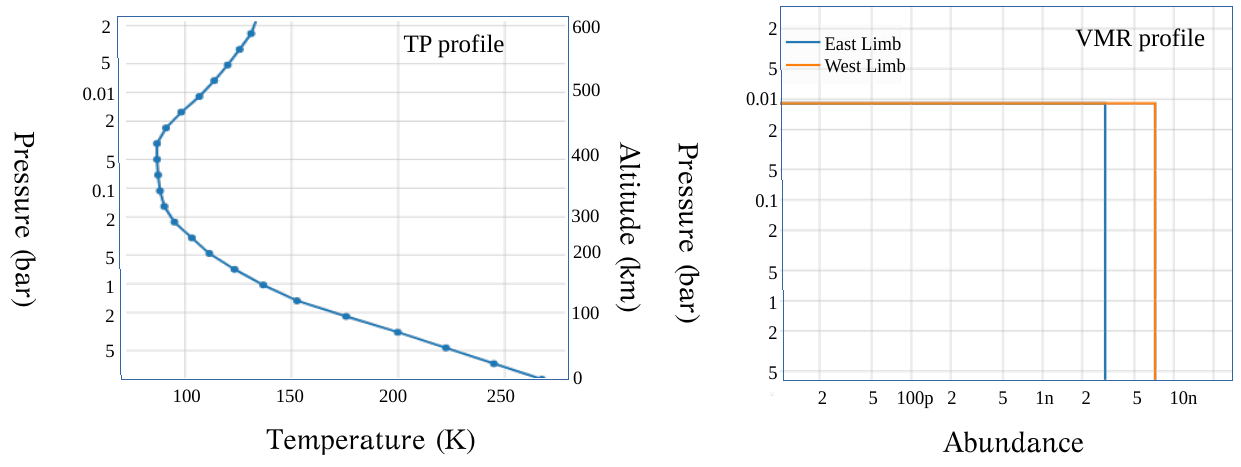}
	\caption{TP and VMR retrieval profiles of HCN in the stratosphere of Saturn using the PSG. In the TP and VMR-retrieved profiles, the pressure level varies from 10 bar to 100 $\mu$bar at altitudes of 0 to 600 km.}
	\label{fig:retrival}
\end{figure*}

\subsection{Continuum emission towards Saturn}
We present the 856.19 $\mu$m (350.14 GHz) wavelength continuum emission image of the gas planet Saturn in Figure~\ref{fig:cont}. The surface brightness colour scale in the continuum image has units of a Jy beam$^{-1}$. In the continuum image, the ring pattern of Saturn is clearly visible at sub-millimeter wavelengths. The synthesized beam size of the continuum image was 0.78$^{\prime\prime}\times$0.61$^{\prime\prime}$. The disk average (except ring) continuum flux density of Saturn is 1.60$\pm$0.35 Jy beam$^{-1}$. We estimate the flux density by drawing a 16$^{\prime\prime}$ circle over the disk part of planet. We do not estimate the flux densities of rings because it is very difficult due to their complex structure. 

%Previously, \cite{dav66} detected the first radio continuum emission from Saturn at wavelength 21.2 cm with a flux density of 0.08$\pm$0.02 Jy, which indicates that the obtained flux density was associated with error owing to the low resolution of the telescope. \cite{dav66} also explained that the synchrotron emission was responsible for the continuum emission at a wavelength of 21.2 cm from Saturn, but the continuum emission at a frequency of 350.14 GHz is mainly coming from thermal emission.

\subsection{Line emission towards Saturn}
From the spectral data cubes, we find that the emission line of HCN with transition J = 4--3, whose rest frequency is 354.505 GHz, originates from both the eastern and western limbs of Saturn. We do not detect any line emissions from the equatorial region of Saturn. We also checked the TGSS, NVSS, and Spitzer sky survey data using the coordinates of Saturn, but we did not observe any source close to the limb. The channel maps of HCN in the atmosphere of Saturn is shown in Figure~{\color{blue}A2}. The channel maps clearly indicates the HCN molecule circulating from the eastern limb to the western limbs of Saturn via the north pole. We extract the molecular emission spectra from the eastern and western limbs of the spectral data cubes of Saturn. For spectral extraction, we created a $2.7^{''}$ diameter circular region in the eastern limb of Saturn centred at RA (J2000) = (18$^{h}$35$^{m}$51.873$^{s}$), Dec (J2000) = (--22$^\circ$19$^{\prime}$27$^{\prime\prime}$.982). We applied the same method for spectral extraction from the western limb. We created a $2.7^{''}$ diameter circular region in the western limb centred at RA (J2000) = (18$^{h}$33$^{m}$53.128$^{s}$), Dec (J2000) = (--22$^\circ$19$^{\prime}$29$^{\prime\prime}$.406). The extracted spectra of HCN from the eastern and western limbs of Saturn are shown in Figure~\ref{fig:SPEC}. In Figure.~\ref{fig:SPEC}, the center position (i.e., 0 km s$^{-1}$) is the theoretical peak position of the HCN (354.505 GHz), but we find that the HCN spectra in the eastern limb (red line) is shifted --9.2 km s$^{-1}$ (354.516 GHz) from the center (i.e., 0 km s$^{-1}$), and for the western limb (black line), the peak of the HCN is shifted to 8.36 km s$^{-1}$ (354.494 GHz). This spectral shift could have been caused by strong stratospheric winds between the eastern and western limbs of Saturn. Earlier, \citet{ca10} also showed the millimeter-wavelength spectral line of Saturn is shifted --9.2 km s$^{-1}$ for the eastern limb line-of-sight and 7.7 km s$^{-1}$ for the western limb line-of-sight using the atmospheric radiative transfer model. Our estimated shift in the HCN emission line for both the eastern and western limbs is similar to that of \citet{ca10}. Except for HCN, we also detected emission lines of CO (J = 3--2) at a frequency of 345 GHz from the eastern and western limbs of Saturn. We did not observe any other complex molecules that are chemically connected to HCN, such as CH$_{2}$NH. We did not present the CO emission line in this paper because the emission line of CO is a known molecule in the stratosphere of Saturn \citep{no86, ca09}.

\begin{figure*}
	\centering
	\includegraphics[width=1.0\textwidth]{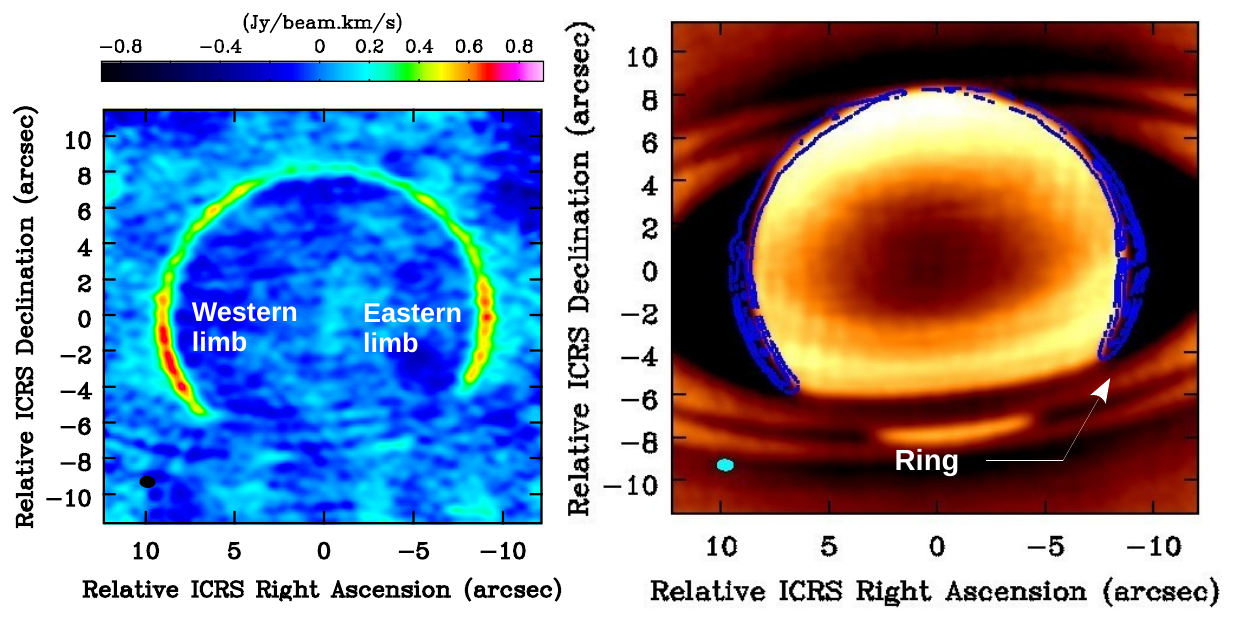}
	\caption{Integrated emission map of HCN in Saturn (left panel). The emission map indicates that the emission line of HCN originates from the eastern and western limbs of Saturn. The right panel image indicates that the HCN map is overplotted with the 856.19 $\mu$m continuum emission map of Saturn. The black and cyan circles indicate the synthesized beam of the integrated emission map and the overplotted map with a size of 0.84$^{\prime\prime}\times$0.61$^{\prime\prime}$. }
	\label{fig:limb}
\end{figure*}

\subsection{Determination of the abundance of HCN}
To determine the abundance of HCN in the atmosphere of Saturn, we used the Planetary Spectrum Generator (PSG, \citet{vil18}). PSG is a radiative transfer suite for spectrum simulation of the solar planets, exoplanets, comets, and other minor objects in our solar system. PSG combines radiative transfer modelling, planetary ephemeris databases (using JPL), and spectroscopic molecular databases to demonstrate the simulated spectra. During the computation of the simulated spectra, we used the line-by-line and correlated-k approaches for better spectral resolution. PSG is also able to fit the modelled spectra of the detected molecules over the observed spectra to determine the abundance of those molecules and the temperature and altitude of the atmospheric levels \citep{vil18}. For spectral modelling, the pressure levels of the vertical atmospheric structure ranged from 10 to 100 $\mu$bar with 60 atmospheric layers. The atmospheric layers are taken from \citet{mos05}. For the temperature profile, we used the disk-averaged retrieved result from \citet{lel84}. We used the HITRAN molecular database for spectroscopic parameters to obtain the abundance of HCN from the western and eastern limbs of Saturn. For fitting the modelled spectra of atmospheric HCN over the observed spectra of HCN, we used the least-squares method, which is well described in \cite{ii20}. During the spectral fitting, the velocity of the western limb spectra is fixed at 8.36 km s$^{-1}$. Similarly, the velocity of the eastern limb spectra is fixed at --9.26 kms$^{-1}$. After spectral fitting, the derived vertical mixing ratios (VMR) of HCN towards the western and eastern limbs of Saturn are 6.19 ppb and 2.90 ppb, respectively. The modelled fitted spectral line of HCN is shown in Figure~\ref{fig:HCNSPEC}. Earlier, \cite{lel84} simulated the VMR in the atmosphere of Saturn to be 3.58 ppb, which is nearly similar to our observed abundance of HCN in the atmosphere of Saturn.

\subsection{Retrieval profile of HCN}
To understand the distribution of HCN in the atmosphere of Saturn, we computed the VMR retrieval profile of HCN using the PSG. We used the optimal estimation method to compute the retrieval model of HCN using the PSG. The pressure levels in the retrieval profile were between 10 bar and 100 $\mu$bar. The temperature profile derived from the disk-averaged result was obtained from \cite{lel84}. For the vertical abundance profiles of CH$_{4}$, C$_{2}$H$_{6}$, C$_{2}$H$_{2}$, NH$_{3}$, and NH$_{4}$SH, except for the HCN, we use default values available in PSG \citep{mos05}. During the VMR analysis of HCN, spectroscopic parameters of HCN are obtained from the HITRAN database. In the VMR retrieval profile of the HCN on Saturn, we used 60 atmospheric layers in the PSG \citep{mos05}. To compute the retrieval profile, we used the abundance of HCN at 6.19 ppb (western limb) and 2.90 ppb (eastern limb) in the PSG. In the retrieval profile, we find that the VMR of HCN is saturated at a pressure level of 0.001 bar. This means that the HCN molecule is distributed up to 475 km in the stratosphere of Saturn. To understand the temperature of the HCN, we also generated the temperature-pressure (TP) retrieval profile of Saturn within a pressure level of 10 bar to 100 $\mu$bar using the PSG. The TP profile clearly shows that the temperature is saturated at a pressure level of 0.001 bar, and within this pressure level, the temperature varies between 100 and 250 K. The TP and VMR retrival profiles of the HCN in the atmosphere of Saturn are shown in Figure~\ref{fig:retrival}.

\subsection{Spatial distribution of HCN in Saturn}
We create an integrated emission map of the HCN in Saturn using the task {\tt IMMOMENTS} in CASA. Integrated emission maps are created by integrating the spectral data cubes in the velocity ranges where the emission line of the HCN is detected. After the generation of the integrated emission map, we find that the emission line of HCN emits from the eastern and western limbs of Saturn. We overplot the integrated emission map of HCN over the 856.19 $\mu$m continuum emission map of Saturn to determine the proper location of HCN emissions in Saturn. The integrated emission map and the overplotted map are shown in Figure~\ref{fig:limb}. We observe that the integrated map of HCN is resolved because the line emitting region is larger than the synthesized beam of the integrated emission map.

\subsection{Possible formation mechanism of HCN in the atmosphere of Saturn}
Earlier, \cite{tok81} claimed that the formation mechanism of HCN in the atmosphere of Saturn is similar to that of Jupiter. \cite{tok81} stated that the photolysis of CH$_{4}$ and NH$_{3}$ is the most likely pathway for the formation of HCN in both Saturn and Jupiter. The laboratory experiment by \cite{rau79} shows that HCN can be produced by photochemical reactions in a mixture of H$_{2}$, CH$_{4}$, and NH$_{3}$ under ultraviolet light. During the experiment, \cite{rau79} observed that NH$_{3}$ abundance is low in the stratosphere of Jupiter, and most of the HCN is produced by photolysis reactions in the lower stratosphere and upper troposphere of Jupiter. If this mechanism holds true in Saturn's atmosphere, the abundance of HCN in Saturn is lower than that in Jupiter because Saturn's atmosphere contains a much lower abundance of NH$_{3}$ \citep{enc74, tok81}. \cite{tok81} calculated the mixing ratio of HCN in Jupiter's atmosphere to be (0.9--4.2)$\times$10$^{-9}$, and we calculated the mixing ratio of HCN in Saturn's atmosphere to be (2.90--6.19) ppb. This comparison indicates that the abundance of HCN in the atmosphere of Saturn is slightly less than the abundance of HCN in the atmosphere of Jupiter. This implies that the laboratory experiment of \cite{rau79} is valid in the atmosphere of Saturn, i.e., the HCN in the atmosphere of Saturn is produced by the photolysis of CH$_{4}$ and NH$_{3}$. Further photochemical modelling is needed to establish the most efficient pathways for the production of HCN in the atmosphere of Saturn. 

\section{Conclusion}
\label{sec:con}
In this paper, we analyzed the ALMA band 7 data for Saturn. The main results are as follows: \\
$\bullet$We successfully detected the rotational emission line of HCN with transition J = 4--3 from the atmosphere of Saturn using ALMA. \\
$\bullet$We observe that the emission line of the HCN mainly arises from the eastern and western limbs of Saturn.\\
$\bullet$ The derived abundances of HCN in the western and eastern limbs are 6.19 ppb and 2.90 ppb, respectively. We also computed the TP and VMR retrieval profiles using the PSG. After retrieval analysis, we observed that the HCN is distributed up to 475 km in the stratosphere of Saturn. Similarly, from the TP profile, the temperature of HCN varied between 100 and 250 K.\\
$\bullet$We also discuss the possible formation mechanism of HCN towards Saturn and claim that the photolysis of CH$_{4}$ and NH$_{3}$ is responsible for the production of HCN.

\section*{ACKNOWLEDGEMENTS}We thank the anonymous reviewers for her/his constructive comments that helped improve the manuscript. AM acknowledges the Swami Vivekananda Merit-cum-Means Scholarship (SVMCM) for financial support for this research. This paper makes use of the following data: \\ADS /JAO.ALMA\#2017.1.00636.S. ALMA is a partnership of ESO, NSF (USA), and NINS (Japan), together with NRC (Canada), MOST and ASIAA (Taiwan), and KASI (Republic of Korea), in cooperation with the Republic of Chile. The Joint ALMA Observatory is operated by ESO, AUI/NRAO, and NAOJ.

\bibliographystyle{aasjournal}
%\bibliography{./literature.bib,added.bib} % if your bibtex file is called example.bib

%\pagebreak
\begin{appendix}    
\begin{figure}
\centering Appendix
	\includegraphics[width=0.5\textwidth]{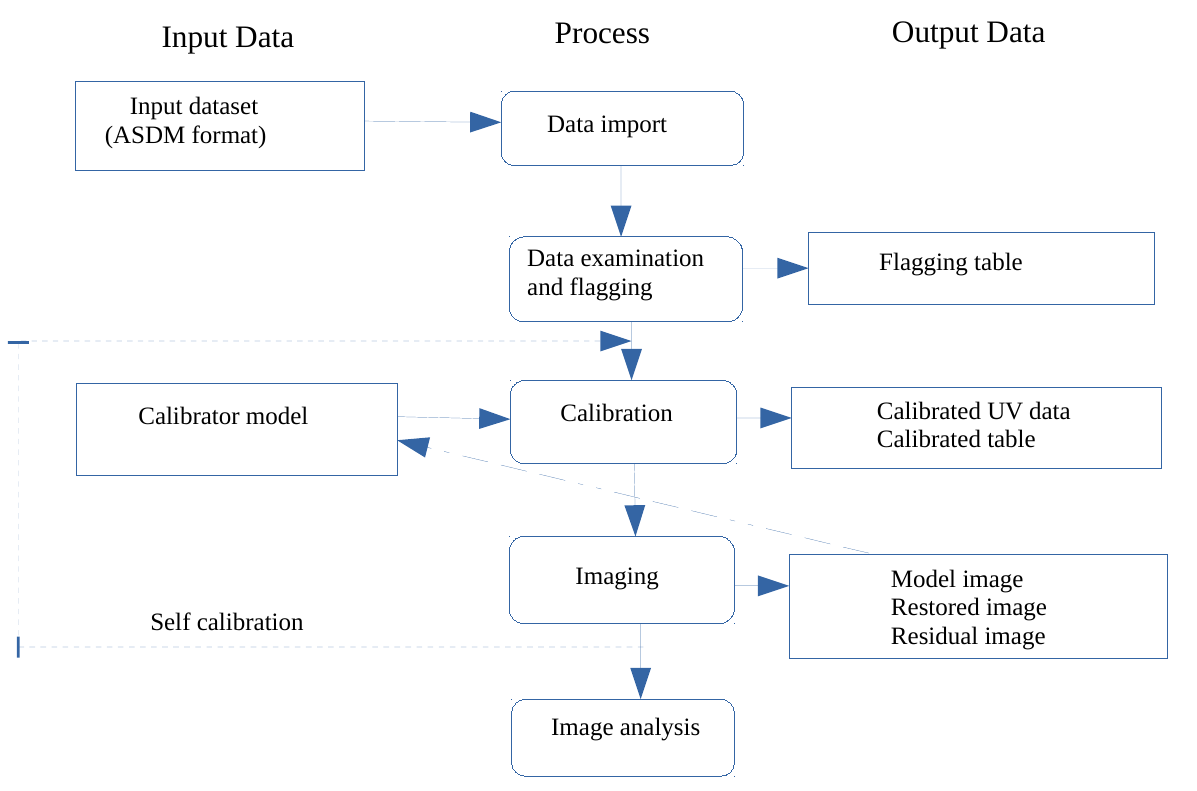}
	\captionsetup{labelformat=empty}
	\caption{Figure A1: An ALMA data analysis flowchart built with the CASA.}
	\label{fig:flowchart}
\end{figure}

\begin{figure*}
	\centering
	\includegraphics[width=1.0\textwidth]{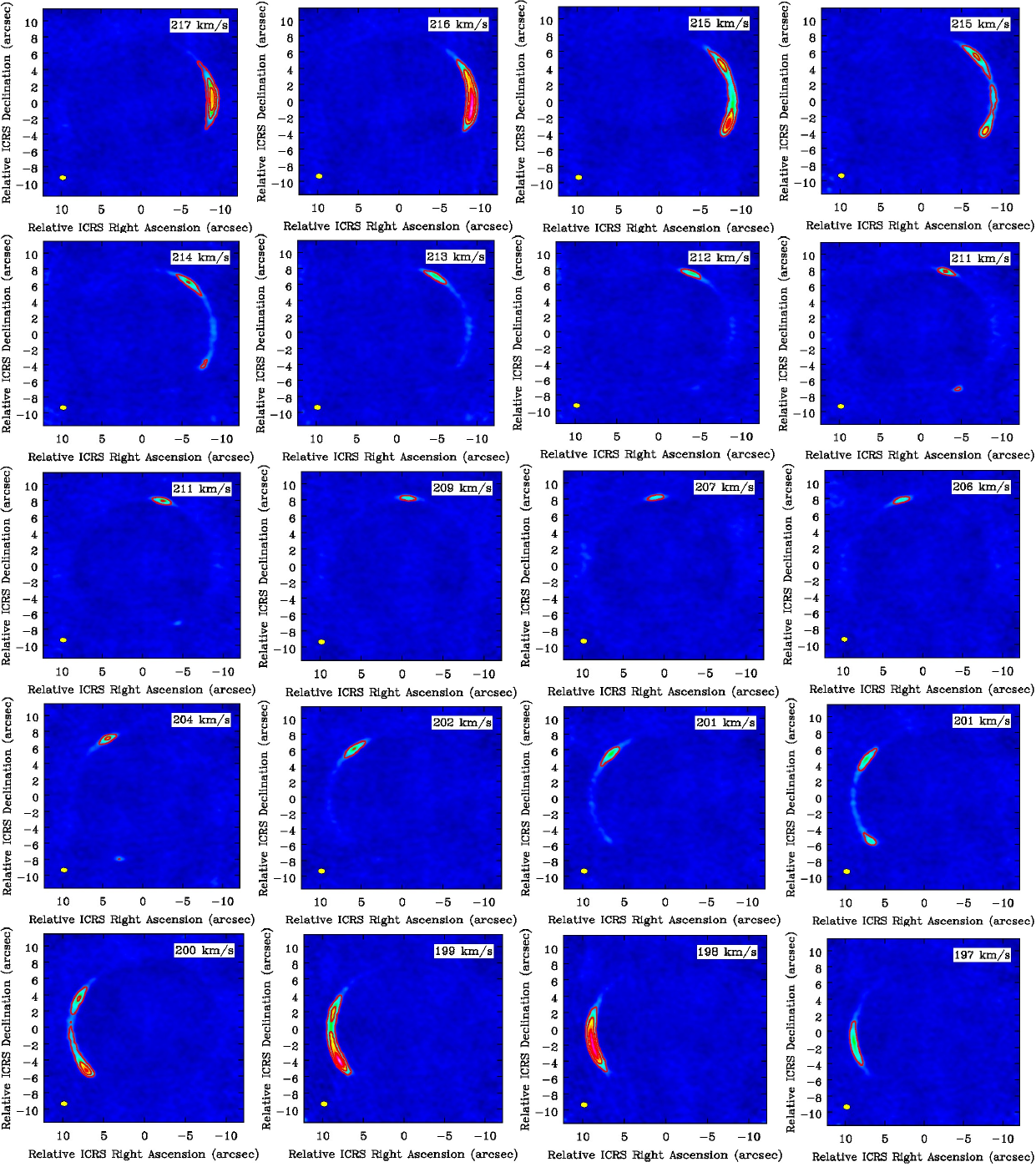}
	\captionsetup{labelformat=empty}
	\caption{Figure A2: Channel maps of the HCN (J = 4--3) in the atmosphere of Saturn. The yellow circles indicate the synthesized beams of the channel maps.}
	\label{fig:linemap}
\end{figure*}
\end{appendix}

\end{document}